\documentclass{rob}%

\usepackage{amssymb}
\usepackage{amsmath}
\usepackage[super]{cite}
\usepackage{algorithm}
\usepackage[noend]{algpseudocode}
\usepackage{xcolor}
\usepackage{tikz}

\newcommand*\circled[1]{\tikz[baseline=(char.base)]{
            \node[shape=circle,draw,inner sep=0.2pt] (char) {#1};}}
            
\newcommand*\squared[1]{\tikz[baseline=(char.base)]{
            \node[shape=rectangle,draw,inner sep=1pt] (char) {#1};}}

\doi{10.1017/xxxx}
\volume{31}
\issue{5}
\pubyear{2013}
\cpyear{2013}
\endpage{7}

\begin{document}

\title[Fabrication-aware Design for Furniture with Planar Pieces]{Fabrication-aware Design for Furniture with Planar Pieces\vspace{1.5em}}

\author{Wenzhong Yan$\dagger$\thanks{Corresponding author. E-mail:
wzyan24@g.ucla.edu}, Dawei Zhao$\ddagger$ and Ankur Mehta\S}
\affil{$\dagger$Mechanical and Aerospace Engineering Department, UCLA, Los Angeles, California\\
$\ddagger$Computer Science Department, UCLA, Los Angeles, California \\
{\S}Electrical and Computer Engineering Department, UCLA, Los Angels, California}

\ADaccepted{MONTH DAY, YEAR. First published online: MONTH DAY, YEAR}

\maketitle

\begin{summary}
We propose a computational design tool to enable casual end-users to easily design, fabricate, and assemble flat-pack furniture with guaranteed manufacturability. Using our system, users select parameterized components from a library and constrain their dimensions. Then they abstractly specify connections among components to define the furniture. Once fabrication specifications (e.g. materials) designated, the mechanical implementation of the furniture is automatically handled by leveraging encoded domain expertise. Afterwards, the system outputs 3D models for visualization and mechanical drawings for fabrication. We demonstrate the validity of our approach by designing, fabricating, and assembling a variety of flat-pack (scaled) furniture on demand.
\end{summary}


\section{Introduction}
Three dimensional (3D) objects built from planar pieces have drawn extensive attention and been wildly applied to owe to their properties, including high strength-to-weight ratio\cite{kim2018origami}, rapid design and  prototyping\cite{yan2018towards}, low cost\cite{yan2020towards}, compact storage and transport\cite{hawkes2010programmable}.
Recently, digital fabrication techniques have greatly increased the ability of casual end-users to create certain physical objects by reducing necessary design and manufacturing investments. However, the creation of functional furniture is still limited to domain experts due to requirements of in-depth engineering understanding for design, skilled carpentry expertise for fabrication and assembly, and material resources to facilitate the whole process. To bring digital fabrication to this space, we have developed a computational design pipeline enabling casual end-users to easily handle the whole creation process of flat-pack furniture, from design, through fabrication, to assembly.

In our system, the design process is abstracted and parameterized, which allows users to easily design their furniture models in a function-based manner without worrying about the low-level engineering implementation. Besides a conventional incremental method, we also harness a hierarchical composition scheme, which further facilitates and accelerates the design process by providing a recursive approach to construct complex models from relatively simple designs through combining functions. Moreover, an intersection auto-detection algorithm is employed to automatically identify connections that are not specified but necessary, and insert planar joints to finalize the designs. In other words, users merely need to specify a minimal number of connections that define the spatial structures of their furniture models; our system will automatically detect other necessary connections and insert specific joints accordingly to generate fabricable embodiment of user-defined models. This algorithm releases users from the tedious connection (and joint) specification process, increasing the flexibility of design process by enabling users to freely build their furniture without being concerned with the order of design. In addition, the incorporation of embedded planar joints within our abstracted and parameterized design scheme greatly reduces the complexity of resulting models and increases the feasibility of assembly for casual end-users. Combining all these features, our system enables casual end-users to easily and intuitively design, fabricate, and assemble flat-pack furniture with guaranteed manufacturability.
\begin{figure}[t]
    \center
    \includegraphics[trim=0cm 23.8cm 0cm 0cm, clip=true,width=1\textwidth]{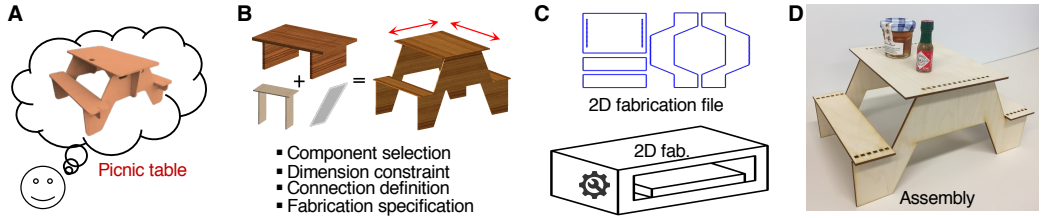}
    \caption{Workflow for making furniture. We use a picnic table as an example. \textbf{A)} Conception of designs; \textbf{B)} Design in our system. Users select components (or designs) from our library, constrains their dimensions, define connections of selected components (or designs), and input fabrication specifications (e.g. materials and corresponding thicknesses); \textbf{C)} 2D fabrication. The output 2D fabrication file is patterned on planar materials (e.g. plywood, 3mm) by 2D fabrication machinery (e.g. laser cutter); \textbf{D)} Assembly. Furniture are built with easy-to-assemble joints through interference fit.}
    \label{fig:designpipeline}
\end{figure}
Using our system, users mainly need three steps to create their desired furniture.  First, designers select components (or use composed designs as components) from our library.  Then they abstractly define dimensions of selected components and specify necessary connections between components to construct furniture models. Our system will automatically handle the detailed fabrication-aware processing to output files for 3D visualization and 2D fabrication.  Finally, users assemble their furniture with ease thanks to the embedded planar joint design. The creation process is simple and intuitive, allowing rapid and easy design of sophisticated furniture for causal end-users. By adopting open-source software, inexpensive raw materials and generalizable fabrication processes, our system expands the accessibility of personalized design to a broader set of non-expert users.  

In this paper, we present the following specific contributions:
\begin{itemize}
 \item a computational design pipeline that allows casual end-users to easily generate customizable, manufacturable, easy-to-assemble flat-pack furniture designs;

 \item an extensible framework that enables users to create furniture designs of arbitrary complexity by hierarchically composing preexisted designs in a function-based manner;
 
 \item an algorithm that greatly increases the feasibility and flexibility of design process through automatically detecting and inserting necessary joints (connections) for user-defined furniture models; and 
 
 \item a representative variety of (scaled) furniture designed and fabricated using the proposed system.

\end{itemize}


 



\section{Related work}
This paper is inspired by the architecture of previous work \cite{Mehta2015, Mehta2014}, which exclusively targets at origami-inspired structures---3D geometries folded from stock sheets of negligible thickness---to create a system which translated structural specifications into a fully functional printable robots. Here, we focus on flat-pack furniture creation with thick sheet materials, which brings in new challenges to design. We also draw upon other academia in the broad categories of modeling by example, fabrication-aware design and personal fabrication. 

\subsection{Modeling by Example}
Shape collections have been widely used to allow data-driven geometric modeling\cite{Xia2016}. Modeling by example\cite{Funkhouser2004,Chen7593384,Schulz2017} enables the users to customize their own models by manipulating existing templates from a large database built by domain experts. More recent work uses recombination of model parts to expand the databases\cite{Kalogerakis2012}. Data-driven suggestions can be used to provide recommendations for designs\cite{Chaudhuri2010}. 

Perhaps the closest work to this project in terms of the desired goals comes from Schulz et. al.\cite{Schulz2014}, in which models---including furniture---could be physically realized via modeling and fabricating by example. From an expert created database of fabricable templates of finished designs, users could manipulate parameterized models with automatically positioning, alignment, and composition. This “model by example” process thus allows casual end users to explore a predefined design space bounded by example designs created by domain experts. However, this “model by example” approach requires extensive efforts from domain experts to build a representative library of the targeted models. In our work, we propose a computational design pipeline for flat-pack furniture. Users are allowed to freely design their desired models with their manufacturability guaranteed. Unlike “model by example” method, no predesigned examples are needed to be created by domain experts in our system; the resulting allowable design space can thus grow unbounded.

\subsection{Fabrication-Aware Design}
Manufacturability of resulting models has long been a concern in the computer graphics community and attracted increasing interest recently\cite{Lau2011}. Fabrication-aware design is proposed to guarantee generating fabricable models with fabrication specifications. It is based on digital parameterization of the building models and then implemented by considering the fabrication specifications through built-in algorithms. These systems with fabrication-aware design aimed at empowering novice users without desired skills to develop real world designs.

Recently, an interactive system SketchChair\cite{Saul2010} was proposed to assist in designing chair models that can easily be fabricated and assembled. This system automatically generates a set of planar pieces that can be intersected along slots to form a 3D realization of a designed chair model. Inspired by the same idea, Chen et. al.\cite{chen2013} and Hildebrand et. al.\cite{Hildebrand}
attempt to convert 3D structures into a set of simplified and fabricable planar polygons connected by interlocking planar pieces. These ideas are developed by building an interactive system where users can have access to real-time feedback by incorporating structure optimization and analysis\cite{Schwartzburg}. Most of that work employs a planar interlock mechanism to implicitly render the design due to its easy manufacturability and assemblability. This interlock mechanism constrains the achievable design space of furniture. Though this method can generate arbitrary solid geometries, it requires dense interactions which may consume excess material.

Our proposed systems makes use of an extensible collection of planar joints to connect different planar elements to generate furniture designs. Similar to the interlocking slots, these joints are rapidly fabricable due to their 2D geometries. A variety of furniture designs have been enabled in this paper. In addition, the design space is potentially extendable thanks to the abstract design scheme. Moreover, other planar joints can be incorporated into our system to enable users to create more different types of furniture.

\subsection{Personal Fabrication and Assembly}
Based on the continued development of democratized tools\cite{Landay2009,Torres2015,Umetani2012,SchulzAdriana2017}, it is expected that casual end-users will play an important role in designing and creating their own products in the future. Several researchers have recently introduced systems for personal fabrication. Spatial Sketch \cite{Willis2010} allows users to sketch in a 3D real-world space. This system realizes the personal fabrication by converting the original 3D sketch into a set of planar slices realized using a laser cutter and assembled into solid objects. Similarly, users can create customized plush toys \cite{Mori2007} and chairs \cite{Saul2010}. These systems convert designs from 3D geometries to a series of planar pieces to simplify fabrication. Post-processing assembly of varying complexity is needed to complete the manufacture. Our work similarly approaches personal fabrication through the use of planar joints which are instinctively easy to assemble, minimizing the need for careful positioning or hardware-based attachments. 

\section{System Overview}
\label{sec:overview}
In this section, we will overview our system by outlining the design workflow and discussing the design space enabled by our computational method. 

\subsection{Design Workflow}
\begin{figure}[t]
    \center
    \includegraphics[trim=0cm 22.4cm 0cm 0cm, clip=true,width=1\textwidth]{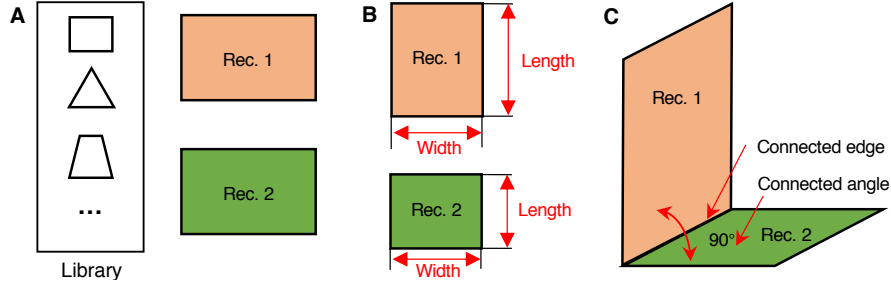}
    \caption{An illustration of a typical design process in our system. \textbf{A)} Users select two predefined rectangle components from our library; \textbf{B)} Users specify the dimension of each component (e.g. widths and lengths of the rectangles); \textbf{C)} Then they define the connection between the two rectangles at selected edges with an 90 degree angle.}
    \label{fig:designprocess}
\end{figure}

As shown in Fig. \ref{fig:designpipeline}, our system assists inexpert end-users in handling the whole creation process of flat-pack furniture, from design, through fabrication, to assembly. More specifically, with a furniture model in mind, user can follow these 4 sequential steps (Fig. \ref{fig:designpipeline}B) to achieve their designs: 1) components {\bf selection}; 2) parameter {\bf constraint}; 3) connection {\bf definition}; and 4) fabrication {\bf specification}. To demonstrate the process, we use a very simple model (two rectangles connected perpendicularly along an edge---perhaps serving as a bookend---as shown in Fig. \ref{fig:designprocess}) as example. In the first step, casual end-users {\bf select} two pre-defined parameterized rectangle components from an existing library (Fig. \ref{fig:designprocess}A) to match their conception of the designs (experts can create their own components, see Section \ref{sec:component}). Then, they {\bf constrain} the selected components' geometries, i.e. widths and lengths (Fig. \ref{fig:designprocess}B). In the third step, users construct the design by {\bf defining} the connections with connected edges and angle (Fig. \ref{fig:designprocess}C). In the last step, given a description of available fabrication {\bf specifications} (e.g. tools, materials), our system will automatically produce manufacturable specifications that capture dimensioned geometries, joint types and joint patterns into a 3D rendering file and a 2D fabrication file. The 3D file can be used to visualize the design. The 2D file can be directly sent to 2D fabrication machines (e.g. laser cutter); thus the resulting fabricated components could be assembled into the desired physical models. Users can repeat step 1-3 to construct designs of arbitrary complexity. In addition, hierarchical composition (see Section\ref{sec:hierarchicalcomp}) could also be harnessed to create sophisticated furniture models. It is worth noting that we represent all components as 2D polygons since they are defined as planar geometries in our system and only substantialized as 3D structures when fabrication specifications (especially material and its thickness) are resolved.

\subsection{Design Space}
Despite the arbitrary complexity of designs allowed in our system, the design space is confined by the physical limitations of the joints. Currently, we employ three types of joints, i.e. finger-finger joint, finger-hole joint and slot-slot joint (as shown in Fig. \ref{fig:joints}A, B, and C), to implement corresponding connections. These joints are instinctively easy to assemble, minimizing the need for careful positioning or hardware-based attachment for casual end-users. However, more types of joints can be implemented to expand the design space to enable more functionalities of resulting models (e.g. active furniture), as demonstrated in Fig. \ref{fig:joints}D-G.

\section{System Implementation Workflow}
In this section, we outline the 5 main steps of system implementation workflow, including parameterized abstraction, hierarchical composition, coordinate placement, intersection auto-detection, and output and assembly. 1) All design elements, including components, connections, and resulting models, are parameterizedly abstracted, which allows users to design furniture in a function-based manner; 2) A hierarchical composition algorithm is then employed to enable users to create complex furniture by functionally combining existing furniture models; 3) the 3D coordinates of all components are computed through a coordinate placement algorithm; 4) these coordinates are then fed into an intersection auto-detection algorithm to automatically identify necessary connections and thus place joints to corresponding positions according to the input fabrication specifications to finalize the models; 5) the resulting furniture models are output with 2D fabrication files (such as .DXF and .SVG) and a 3D .STL file, ready for fabrication and assembly. In addition, we describe the main operational features of our system in detail. 

\subsection{Parameterized Abstraction}
The 3D geometries of flat-pack furniture in our system are defined as a composition of components connected with some connections. In traditional furniture design process, creating a piece of functional furniture design can be rather challenging since users may need to adjust many parameters with complex dependencies while maintaining the manufacturability and assemblability. Even with the aid of some CAD tools, users may still experience great difficulties due to the lack of in-depth understanding of CAD software and the manufacturing process. 

In order to allow casual end-users to design furniture with ease, our system largely simplifies the design process into a abstracted function-based manner. To achieve this abstracted design, we parameterize all elements in furniture creation process, including components, connections, designated furniture models, and fabrication specifications. Therefore, we introduce the parameterized abstraction of components, connections, and furniture models in following paragraphs.

\subsubsection{Components}
\label{sec:component}
\paragraph{\textbf{Representation}}
As mentioned before, components are represented as 2D polygons. Therefore, in our system, a component's fabrication-related parameters, e.g. material type and its thickness, are not defined until specific fabrication and assembly methods are determined by users in the final steps of the design process. At that time, these abstract components are implemented automatically as 3D ingredients according to the input fabrication specifications. 

Components in our system fall into two categories: basic components and hierarchically composed components (later introduced in Section \ref{sec:hierarchicalcomp}). Our system already comes with a set of predefined, commonly-used polygon components, such as rectangle, trapezoid, n-side polygon, etc. After abstract parameterization, every component in our system is programmed as an object instantiated from a corresponding component class, and represented as follows:
$$ \textbf{Component}\;(para\,\textbf{1}, para\,\textbf{2}, ...,para\,\textbf{i}, ..., para\,\textbf{n}) $$
where \textbf{Component} is the component class name and $para\,\textbf{i}$ is the i\textsuperscript{th} predefined parameter of the component. To instantiate a component object of the corresponding component class, users only need to specify the parameters.

\begin{figure}[t]
    \center
    \includegraphics[trim=0cm 24.5cm 0cm 0cm, clip=true,width=1\textwidth]{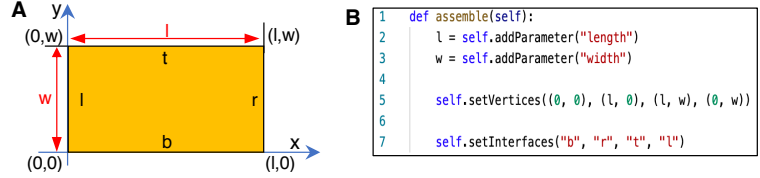}
    \caption{An example of parameterized abstraction. \textbf{A)} Rectangle component geometric diagram with parameters labeled; \textbf{B)} Program implementation of a parameterized rectangle class in Python script in our system.}
    \label{fig:component}
\end{figure}

\paragraph{\textbf{Construction}}
Fig. \ref{fig:component} uses a rectangle component as an example to demonstrate how we construct a component class in our system (each component is an instance of the component class). Firstly, users need to define the parameters of the component class (line 2-3), which are the length $l$ and width $w$ of the rectangle in this case. Then, users will specify the outer vertices of the component class whose order follows the right-hand rule so that the front of the component is facing the out-of-page direction (line 5). Lastly, users set a series of interfaces to the component class, which are some ports that can be used to connect with the interface of other components. In this example, we set the edges of the rectangle as interfaces with name $b$, $r$, $t$, and $l$. Thus, we have a rectangle class with 4 interfaces and two parameters. This mechanism applies to other component classes.
 
Users can also define their own custom components following the same workflow illustrated above, which only requires them to specify the parameters, vertices, and interfaces of the components. Customizing components is one way to define desired components when they are not available in our library. Also, users can harness hierarchical composition (see Section \ref{sec:hierarchicalcomp}) to piece together elementary components into new complex component. For example, we can obtain a “L” shape new component by combining two rectangles together along edges. It is worth noting that users are allowed to define components with fewer parameters, called meta-parameters, by adding some geometric constraints to the original parameters. Thus, users won't be exposed to enormous design parameters when design becomes complicated. For example, we can only select the length of the rectangle as the manipulable meta-parameters and geometrically constrain its width as a half of its length to create a component with a fixed aspect ratio.

\subsubsection{Connection}
\begin{figure}[t]
    \center
    \includegraphics[trim=0cm 23.1cm 0cm 0cm, clip=true,width=1\textwidth]{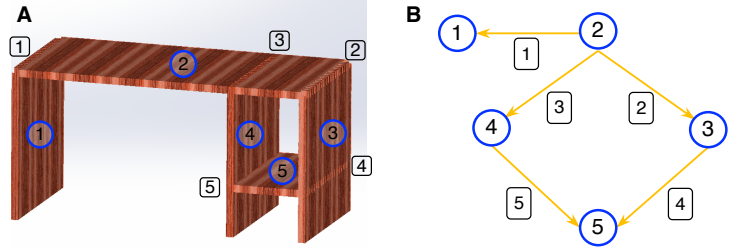}
    \caption{Representation of a furniture model. We take a computer desk as an example. Components labelled as circled number, e.g. \circled{1}) and connections as boxed number, e.g. \squared{1}. \textbf{A)} 3D model of the desk; \textbf{B)} Connectivity graph of the desk with connections represented as directed yellow lines, whose directions indicate the connection orientation.)}
    \label{fig:connectionrepr}
\end{figure}

\paragraph{\textbf{Representation and Construction}} Connections in our system are also parameterizedly abstracted. Once a connection has been defined, we build an associated connectivity item to store all its information that is efficient to embody the connection physically in following operations. A connection is represented by a directed line in a connectivity graph of a furniture model. Here we use a computer, as shown in Fig. \ref{fig:connectionrepr} as example. In the figure, connection \squared{1} ($\circled{2}\xrightarrow{}\circled{1}$) denotes that component \circled{2} is connected to component \circled{1} at some interfaces. In the graph, the edges of the line represent the connected interfaces of corresponding components. The direction specifies the way we construct the connection. Note that \squared{1} ($\circled{1}\xrightarrow{}\circled{2}$) is different from \squared{1} ($\circled{2}\xrightarrow{}\circled{1}$). More explanations would be found in the following paragraph.

Finding a proper way to specify the abstracted connection between two components can be non-trivial because it needs to be accurate, concise, and intuitive for users, while being able to express all possible spatial relations. In our system, the connection is constructed abstractly as a 5-tuple. For instance, when component $C_A$ is connected to $C_B$, we can call the constructor to define the connection:
$$\textbf{Connection}(\:(\textbf{$C_A$}, I_{A}^i), (\textbf{$C_B$}, I_{B}^j), P_{A}, P_{O}, P_{R})$$
\noindent where $C_A$ and $C_B$ are two connected components. In this expression, the first component $C_A$ is connected to the second component $C_B$. In this connection, $C_A$ is named as the connecting component and $C_B$ is named as the connected component. Thus, this connection could be shortly annotated as $\textbf{Connection}(A \xrightarrow{}\textbf{B)}$. The sequence of the two components matters. $I_{A}^i$ and $I_{B}^j$ represent the selected interface $i$ of component $C_A$ and the interface $j$ of component $C_B$. These two selected interfaces will be connected together. $P_{A}$, $P_{O}$, $P_{R}$ represent the alignment, offset, and rotation of the connection. The value of $P_{A}$ is either “front-front” or “front-back”. “Front-front” means that two connected components have the same orientation while “front-back” indicates their orientation is opposite to each other. $P_{O}$ is a 3-tuple, specifying the 3D offset vector that component A (connecting component) should travel in accordance with. $P_{R}$ is also a 3-tuple, defining how component A along with its local coordinate system is rotated around its own x, y, and z axis, respectively. For example, in Fig. \ref{fig:connectionrepr}B, the connection between component \circled{1} and \circled{2} can also be described as $\textbf{Connection}(\circled{2} \xrightarrow{} \circled{1})$.

\begin{figure}[t]
    \center
    \includegraphics[trim=0cm 16.2cm 0cm 0cm, clip=true,width=1\textwidth]{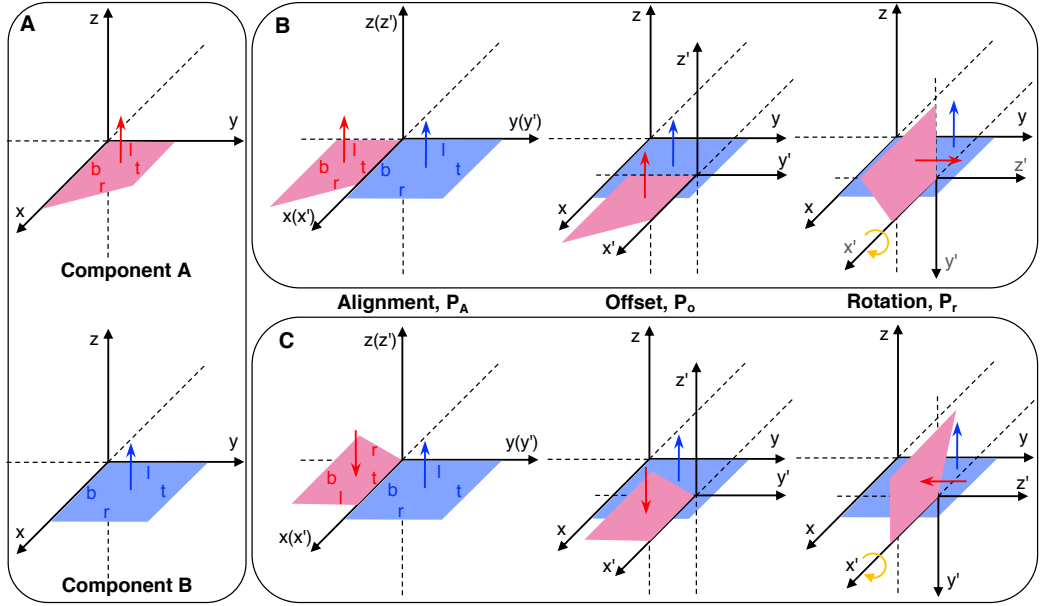}
    \caption{Connection visualization. \textbf{A)} Component A and B with their interfaces labelled and original coordinates specified; Connections with “front-front” alignment \textbf{B)} and “front-back” alignment \textbf{C)}. Left: Alignment defined; Middle: 3D offset; Right: 3-axis rotation.}
    \label{fig:connectionrule}
\end{figure}

Using this 5-argument connection constructor, we can represent any possible spatial relations between two components while providing users an intuitive and easy way to state connections. Here we will illustrate this connection constructor in detail (see Fig. \ref{fig:connectionrule}). For instance, $\textbf{Connection}(\:(\textbf{$A$}, t), (\textbf{$B$}, b), front-front, (v_x,v_y,v_z), (\theta_x,\theta_y,\theta_z))$ can be visualized as in Fig. \ref{fig:connectionrule}B. From the connection constructor, we know that interface $t$ of component A is connected to interface $b$ of component B. Since the alignment is specified as “front-front”, the orientation of both faces facing up results in a temporary position as the left graph in Fig.\ref{fig:connectionrule}B. x-y-z is the coordinate of component B being set as the global coordinate for this connection while x'-y'-z' is the local coordinate of component A, fully overlapping with B's. Then a 3D offset vector $(v_x,v_y,v_z)$ is applied to component A (as the middle graph shows) and followed by a 3-Axis rotation $(\theta_x,\theta_y,\theta_z))$ to change the orientation of the component (e.g. (90,0,0) represents a 90\textsuperscript{$\circ$} rotation about x axis as in the right graph in Fig. \ref{fig:connectionrule}B). The other case with “front-back” alignment is also presented in Fig. \ref{fig:connectionrule}C. Though rather simple and intuitive, the connection specification process could be greatly simplified with a graphic user interface later on.

\paragraph{\textbf{Physical Implementation}}
Though parameterized abstraction can greatly facilitate the design process, the connections need to be implemented physically to actually join connected components when it comes to manufacture. In this paper we decide to use planar joints to embody the connections to reduce the difficulty of fabrication and assembly. Thus, users do not need to go through the tedious assembly process with system-defined joints \cite{Schulz2014}.

\begin{figure}[t]
    \center
    \includegraphics[trim=0cm 18cm 0cm 0cm, clip=true,width=1\textwidth]{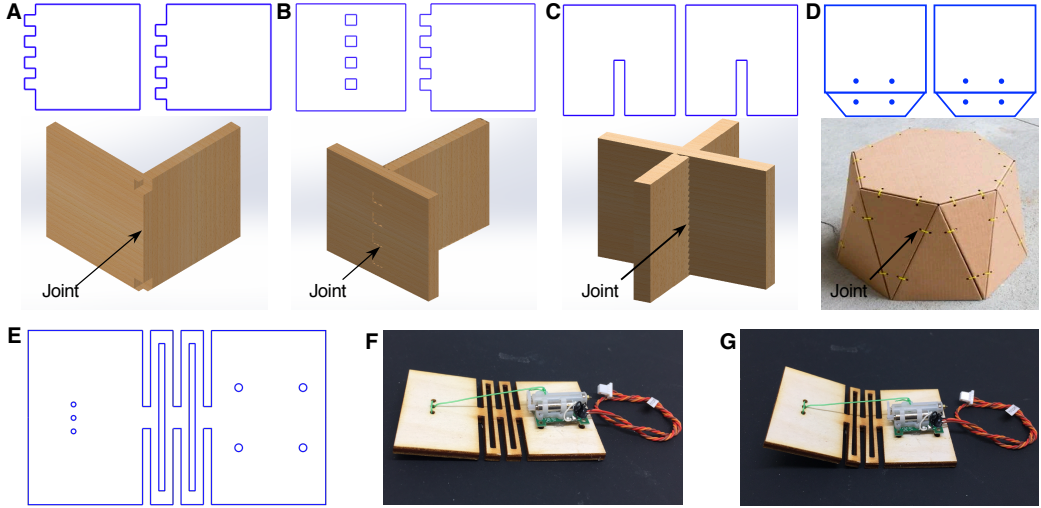}
    \caption{Joint collection. \textbf{A)} A finger-finger joint for an edge-edge connection; \textbf{B)} A finger-hole joint for an edge-face connection; \textbf{C)} A slot-slot joint for a face-face connection; \textbf{D)} A flap joint for edge-edge connection\cite{joinery2018}. \textbf{E} The fabrication pattern of a cable-driven joint. The relax and bent states of the joint are shown in \textbf{F} and \textbf{G}, respectively. It is worth noting that cables are needed for both flap joints and cable-driven joints and linear motors are required for cable-driven joints.}
    \label{fig:joints}
\end{figure}

To enable the manufacturability of furniture designs, joints must be added at places of intersections. However, for casual end-users, determining and drawing proper joints patterns can be a liability, even with the help of some design software like AutoCAD or Inkscape. Our computational design tool will automatically add one of the three types of joints (as shown in Fig. \ref{fig:joints}A, B, and C, respectively) according to specific abstract connections. More details for these three types of joints can be found in Appendix \ref{sec:joints}). The joints that we introduce here have the following two advantages: 1) they can be easily fabricated using modern 2D manufacturing tools (e.g. laser cutter, waterjet, and jigsaw); 2) are handy to assemble even without skilled craftsmanship.

Theoretically, any joints, especially those that satisfy the above requirements, can be incorporated into our system since the joints are merely the physical implementation of abstracted connections\cite{Tian2018}. For example, the flap joints\cite{joinery2018} (see Fig.\ref{fig:joints}D, not implemented in our system) could potentially be adopted to achieve edge-edge connections. Further work could be done to extend our library of joints to expand the design space. 

In addition, active joints could be possible to realize connection implementation, which leads to active devices, such as active furniture and robots. As a proof of concept, we create cable-driven joints to allow resulting devices to have angular movements. For example, we connect two rectangle components with a cable-driven joint, as shown in Fig.\ref{fig:joints}F and G. The fabrication pattern is shown in Fig.\ref{fig:joints}E. By adding a lattice pattern \cite{Blashki2015} (more details can be found in Section \ref{sec:flexiblejoint}) along the connected edges of the two rectangles, a flexible joint is formed to allow angular movements between two connected components. A linear motor with its shaft is tied to the left rectangle using a cable, is then attached on the right rectangle, as shown in Fig.\ref{fig:joints}F. The back-and-forth movement of the shaft of the motor will change the angle between two components (see Fig.\ref{fig:joints}F and G). Presumably, this type of joints can be used as living hinges for doors of furniture or movable joints of robots. 

In this paper, joints are mainly assembled through interference fit, which means optimal interference amplitudes are needed to be calculated based on the fabrication specifications. Our system can automatically output the optimal profiles of joints once the fabrication specifications (e.g. materials and fabrication machines) are defined by users. More details could be found in Appendix \ref{sec:joints}.

\subsubsection{Model}
\hfill \break
Using the connection constructor specified on relevant components, an abstract design model is built internally to store all the information (including components with specific geometry constraints and connections between components) relevant to the design. This designed model can be visualized as a directed connectivity graph. Here, we use a computer desk as an example. As shown in Fig. \ref{fig:connectionrepr}A, the computer desk consists of five rectangle components and five connections. The connectivity graph of the computer desk is shown in Fig. \ref{fig:connectionrepr}B. Each component is labeled as a circled number, such as \circled{1} and each connection is represented by a numbered symbol, e.g. \squared{1}. Each directed edge denotes a connection with its vertices intersected with circles at their corresponding interfaces. For example, component \circled{1} is connected to one interface of component \circled{2} resulting in connection \squared{1} while component \circled{3} is connected to another interface of component \circled{2} to form connection \squared{2}. Thus, a connectivity graph effectively includes all information relevant to the corresponding design. By traversing the graph following the algorithm presented in Section \ref{sec:PnP}, we can generate the 3D coordinates of each component of the design for further operations. Each connectivity graph, representing a design, can also be stored in our library as a .YAML file, which can be in turn used as a new component to hierarchically compose more complex designs in a function-based manner.

\subsection{Hierarchical Composition}
\label{sec:hierarchicalcomp}
\subsubsection{Design Principle}
\begin{figure}[t]
    \center
    \includegraphics[trim=0cm 20.3cm 0cm 0cm, clip=true,width=1\textwidth]{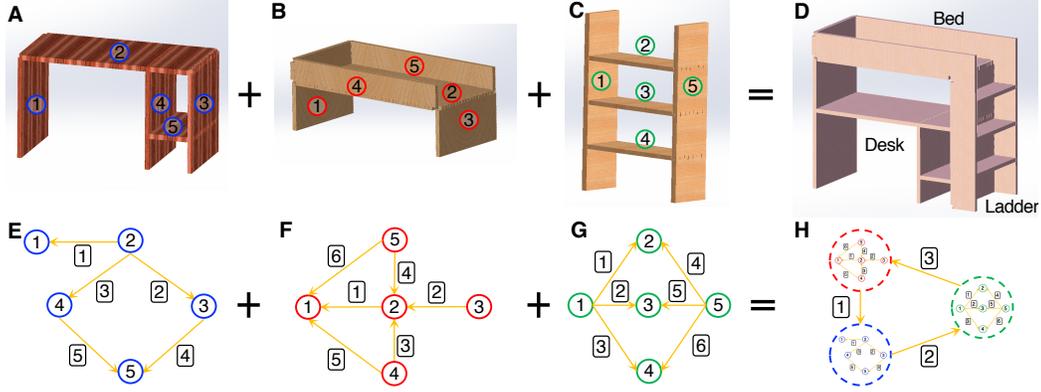}
    \caption{A illustration of hierarchical composition with a bunk bed. Three “components”, i.e. a computer desk \textbf{A)}, a bed \textbf{B)}, and a ladder \textbf{C)} are composed into a bunk bed \textbf{D)}. Each “component” itself is a furniture design with certain functionalities. The connectivity graphs of three “components” are also pictured in \textbf{E)}, \textbf{F)}, and \textbf{G)}. The final connectivity graph of the bunk bed is also directly composed of all “components” \textbf{H)}.}
    \label{fig:hierarchicalcomp}
\end{figure}

\hfill \break
To enable users to build complex furniture designs with ease, we harness function-based hierarchical composition, which allows users to recursively build up to the desired complex furniture constructions from relatively simple existing designs. This means we organize all the parameterized data of the furniture design into a hierarchical tree, whose structure is defined by how we recursively create the furniture. In this manner, functional models at any level could be treated as components to construct higher level furniture designs. For instance, to build a bunk bed as illustrated in Fig. \ref{fig:hierarchicalcomp}D, instead of building the whole piece from scratch, user can simply select these existing designs, a computer desk, a bed, and a ladder (as shown in Fig. \ref{sec:hierarchicalcomp}A, B, and C, respectively) to be composed together. The computer desk, bed, and ladder can be further decomposed into elementary components (e.g. rectangle). Therefore, the $i_{th}$ level hierarchy $M^{i}$ of a hierarchically composed model can be written as:
$$M^{i} = (\{Component^i\}, \{Connection^i\}, \{Constraint^i\}, \{Interface^i\})$$
\noindent where $\{Component^i\}$ is a set of “basic” components consisting of the hierarchical design in this level. Each “basic” components in this set can be a composed furniture design or an elementary component. $\{Connection^i\}$ is a set of connections specifying how the “basic” components are connected, $\{Constraint^i\}$ is the set of parameters constraining model $M^{i}$, and $\{Interface^i\}$ is the set of interfaces of the new hierarchically composed model $M^{i}$.

\subsubsection{Model Reconstruction}
\hfill \break
Each design in our system is represented and stored as an abstract connectivity graph. When users compose several simpler designs into a more complex functional model, all connectivity graphs will be added to an new high-level connectivity graph as per the connections specified at this hierarchy by the user. This means that all information of each simpler design is integrated to construct a new hierarchical data tree with its own components and connections preserved, as shown in Fig. \ref{fig:hierarchicalcomp}E-H. In the same manner, this hierarchical data tree will be integrated as a branch of a new tree at a higher hierarchy when current design is composed into a more complicated furniture.

\subsection{Coordinate Placement}
\label{sec:PnP}
To place each component in a global coordinate system and find the places of intersections to add proper joints, we employ a coordinate placement algorithm. At the design stage, we require users to define connections between components. However, these connections only specify the relative spatial relation, which is not necessarily the positions at which joints are placed. Therefore, we use an intersection auto-detection algorithm (see Section \ref{sec:intersectiondete}) to place necessary joints based on the derived spatial relation to finalize the design for manufacturing and assembly. Having derived the connectivity graph representing the furniture designs, our system can perform a traversal on the graph to recursively compute the 3D coordinates of each component following Algorithm \ref{alg:3D}, which will later be used to build up the 3D model.

\begin{algorithm}[t]
\caption{Compute the global 3D coordinates for every component of the component set \{C\} of a design based on the defined associated connections}
\label{alg:3D}
\begin{algorithmic}[1]
\State Randomly select a component $C_i$ from \{C\}  \hfill\break
\textcolor{red}{\hspace{0.0cm}//Return coordinates of components connected to $C_i$}
\Function{Find3D}{$C_i$, $T_{global}$}
    \State $C_i .T_{3D} \gets T_{global}$
    \For {\textbf{Connection}$(A\xrightarrow{}\textbf{B)}$ in $C_i.Connections$} \hfill\break
        \textcolor{red}{\hspace{1cm}//$C_i$ is connected component}
        \If{$C_i \: is \: C_{B}$}
            \State $T_{relative} \gets FindRT(C_{A \gets B})$
            \algorithmiccomment{Find relative transformation matrix}
            \State $T_{global}^{'} \gets T_{global} \cdot T_{relative} $
            \algorithmiccomment{Calculate new global transformation matrix}
            \State return Find3D($C_{A}$, $T_{global}^{'}$) \hfill\break
        \textcolor{red}{\hspace{1cm}//$C_i$ is connecting component}
        \ElsIf{$C_i \: is \: C_A$}
            \State $T_{relative} \gets FindRT(C_{A \gets B})$
            \algorithmiccomment{Find relative transformation matrix}
            \State $T_{global}^{'} \gets T_{global} \cdot T_{relative}^{-1} $
            \algorithmiccomment{Calculate new global transformation matrix}
            \State return Find3D($C_B$, $T_{global}^{'}$)
        \EndIf
    \EndFor
    
\EndFunction

\end{algorithmic}
\end{algorithm}

Initially, each component is placed within its own local coordinates as defined by the user as in Fig. \ref{fig:component}A. Our goal is to find each component's 4 by 4 transformation matrix, $T_{global}$, which transforms the homogeneous coordinates of the original components into a global 3D coordinates space as per the connections. To do so, we adopt the recursive Algorithm \ref{alg:3D}. The input of the function consists of the component set $C$, a 4 by 4 transformation matrix $T_{global}$ that represent its global coordinate. The algorithm randomly selects a component $C_i$ as a starting point and recursively find the coordinate of all components connected to it. This step starts by storing the transformation matrix as an attribute of the component. Then, for each connection, e.g. $\textbf{Connection}(A\xrightarrow{}B)$, that involves this component, if this component $C_i$ is connected to component $C^B$ (i.e. $C_i=C_B$, line 5), we can find the relative transformation, $T_{relative}$, through the function, $FindRT(\textbf{Connection}(A\xrightarrow{}B))$, between them and times the global transformation matrix of $C_B$ (= $C_i$) to obtain the new global transformation matrix $T_{global}^{'}$ to run the next iteration until the 3D coordinates of all components are calculated (line 6-8). Another execution will be made for $C_A$. If $C_i$ is connecting component $C^A$ (i.e. $C_i=C_A$, line 9), then we should find the inverse of the relative transformation between them and times the global transformation matrix of $C_A$ (= $C_i$) to have new global transformation matrix $T_{global}^{'}$ to run the next iteration (line 10-12).

\subsection{Intersection Auto-Detection}
\label{sec:intersectiondete}
After the coordinate placement phase illustrated in Section \ref{sec:PnP}, we derive the global 3D coordinates for all components, but the design is still by no means manufacturable. In other words, we need to detect the places of intersections and add proper joint mechanisms to finalize the model as well as to ensure its manufacturability. By performing an automatic intersection detection, we release users from the burden of specifying all places of intersections with connections explicitly. More specifically, using the connection constructor, user can easily build the 3D model for their designs with a minimal number of connections, which may not include all necessary intersections. Then our algorithm can automatically handle the intersection detection to guarantee the manufacturability and functionality. This algorithm is particularly useful for hierarchical composition because it is very challenging for causal end-users to specify all necessary connections thoroughly due to the geometry complexity. Hence, it help users to focus on design in a function-based manner.
\begin{figure}[t]
    \center
    \includegraphics[trim=0cm 22.1cm 0cm 0cm, clip=true,width=1\textwidth]{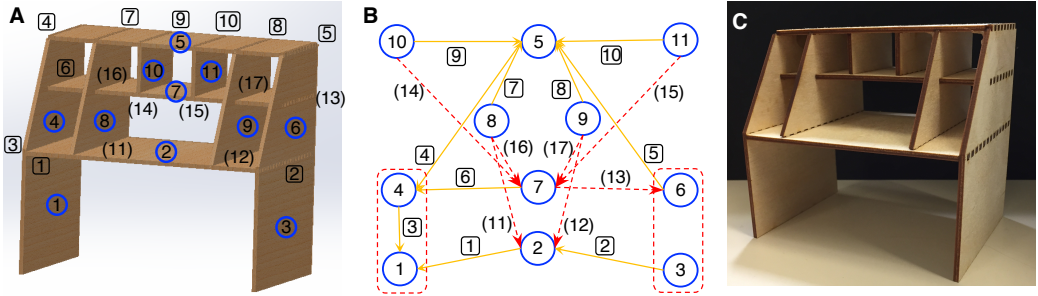}
    \caption{An illustration of intersection auto-detection. Components labelled as circled number, e.g. \circled{1}) and connections as boxed number, e.g. \squared{1}. \textbf{A)} 3D model of a reading desk; \textbf{B)} Connectivity graph; \textbf{C)} Fabricated and assembled (scaled) reading desk with 3mm plywood. Note: this desk can also be built through hierarchical composition, which will be discussed later in Section \ref{sec:modelingFlexibility}.}
    \label{fig:connectiondetec}
\end{figure}

In this section, we use a reading desk, as shown in Fig. \ref{fig:connectiondetec}A, as an example to illustrate the algorithm. To design such a reading desk with 11 components and 17 joints, user only need to specify 10 connections (from \squared{1} to \squared{10} in Fig. \ref{fig:connectiondetec}B) with other necessary intersections (or joints) (from (11) to (17)) inserted automatically by our algorithm. For instance, a joint is automatically added between component \circled{2} and \circled{8} while there is no connection defined by users. Moreover, our algorithm can help to merge coplanar components to reduce the complexity of design and assembly. For example, component \circled{1} and \circled{4}, originally connected by connection \squared{3}, are merged into a single component.

In the following two sections, we will explain the intersection auto-detection algorithm in detail with which divided into two parts: coplanar faces merging (line 1-6), intersection segments searching (line 7-16), and joints inserting (line 17-27).
\begin{algorithm}[t]
\caption{Find all necessary intersections and joints of a designed model composed of a set of components \{C\} and specified connections}
\label{alg:intersection}
\begin{algorithmic}[1]
\Function{IntersectionDetection}{$ \{C\} $}\hfill\break
    \textcolor{red}{\hspace{0.54cm}//Merge coplanar faces}
    \For {$C_i$ in $\{ C \}$}
        \For {$C_j$ in $\{ C \}$}
            \If {$coplanar(C_i, C_j) \land C_i \cap C_j \neq \emptyset $}
                \State $V_{i \cup j} \gets (C_i\cup C_j)$ 
                \algorithmiccomment{find the geometric union of $C_i$, $C_j$}
                \State $C_i.V^{2D} \gets {C_i.T^{3D}}^{-1} \cdot V_{i \cup j}$ 
                \algorithmiccomment{convert to global coordinate system}
                \State $C_j.V^{2D} \gets \emptyset $
            \EndIf
        \EndFor
    \EndFor
    \break
    \textcolor{red}{\hspace{0.54cm}//Find segments of intersections of each component}
    \For {$C_i$ in $\{ C \}$}
        \For {$C_j$ in $\{ C \}$}
            \If {$ \neg coplanar(C_i, C_j) \land \neg parallel(C_i, C_j) $} 
                \State $l \gets C_i \cap C_j$
                \algorithmiccomment{find the line of intersection between $C_i$ and $C_j$}
                \State $\{s_i\} \gets l \cap C_i$
                \algorithmiccomment{find segments of intersection between $l$ and $C_i$}
                \State $\{s_j\} \gets l \cap C_j$
                \algorithmiccomment{find segments of intersection between $l$ and $C_j$}
                \For {$s_k$ in $\{ s_i \} \cap \{ s_j \}$}
                    \State $s_k^i \gets C_i .{T^{3D}}^{-1} \cdot s_k$
                    \algorithmiccomment{convert to global coordinate system}
                    \State $s_k^j \gets C_j .{T^{3D}}^{-1} \cdot s_k$
                    \algorithmiccomment{convert to global coordinate system}\break
                    \textcolor{red}{\hspace{2.82cm}//Add joints according to the types of intersections}
                    \If{$l_i \in edge(C_i) \land l_i \in edge(C_j) $}
                        \State $C_i .finger \gets C_i .finger \cup \{l_i^1 \}$
                        \State $C_j .finger \gets C_j .finger \cup \{l_i^2 \}$
                    \ElsIf {$l_i \in edge(C_i)$}
                        \State $C_i .finger \gets C_i .finger \cup \{l_i^1 \}$
                        \State $C_j .hole \gets C_j .hole \cup \{l_i^2 \}$
                    \ElsIf {$S_i \in edge(C_j)$}
                        \State $C_i .hole \gets C_i .hole \cup \{l_i^1 \}$
                        \State $C_j .finger \gets C_j .finger \cup \{l_i^2 \}$
                    \Else
                        \State $C_i .slot \gets C_i .slot \cup \{l_i^1 \}$
                        \State $C_j .slot \gets C_j .slot \cup \{l_i^2 \}$
                    \EndIf
                \EndFor
            \EndIf
        \EndFor
    \EndFor
\EndFunction
\end{algorithmic}
\end{algorithm}

\subsubsection{Coplanar Faces Merging}
\hfill \break
Frequently, several different components of a furniture design end up being coplanar (or overlapping). Merging them often not only simplifies the design but lowers the difficulty of fabrication and assembly. For coplanar faces merging, we iteratively select two components $C_i$, $C_j$ from the components set $\{C\}$ (line 4) and determine if they are coplanar faces with intersections, which is fairly easy as we have already found the 3D coordinates of all component. Then, we use a geometry boolean operation library to find the union of the 3D global coordinates of $C_i$, $C_j$ (line 5). After finding the union of 3D global coordinates, we need to transform them into the local 2D coordinate system of the planar component so that we can later work on the 2D output file for the design. Therefore, we will dot product the union of the 3D global coordinates with the inverse of the 3D transformation matrix of one component $C_i$ (line 6) and set the vertices of the other component $C_j$ (line 7) to be empty. This means that all information of component $C_j$ is transferred into component $C_i$.

\subsubsection{Intersection Segments Searching}
\hfill \break
For any two components $C_i$, $C_j$ that are not coplanar or parallel, we will first find the intersection line $l$ of the two 3D planes that these components are lying in (line 11). Then, we find the two sets of segments where the intersection line intersects with component $C_i$, $C_j$ (line 12-13). After finding the union of these two line segment sets (line 14), we will also perform coordinate system transform step, similar to line 6, to transform the coordinates from the 3D global coordinate system to the local 2D coordinate system of $C_i$ and $C_j$ (line 15-16). So far, all of the intersection segments within a certain design are identified, ready for joints to be inserted in next step.

\subsubsection{Joints Inserting}
\hfill \break
The last step is to determine the types of joints for each component based on the positions of intersection segments. If the intersection segment of a connection is on the edge of both connected components, then a finger-finger joint will be added on them (line 17-19). If the intersection segment is on the edge for one component and within the face of the other, a finger-hole joint will be selected. Fingers will be added on the former component, and holes will be added on the latter (line 20-25). If the intersection segment is within the face of both components, then a slot-slot joint (line 26-28) will be added on them. All inserted patterns of joints are automatically calculated by considering fabrication specifications (more details about the pattern generating can be found in Appendix \ref{sec:joints}).

\subsection{Output and Assembly}
\label{sec:output}
Each furniture design in this system is an instance of a common parent class, stored as an executable Python script. As mentioned above, in our system, each component is also assigned a set of manufacturing specifications that dictate how the compiler should translate the design specifications to real world output. At execution time, system calls a function to place all the components into a final 2D design file (.SVG or .DXF), as shown in Fig. \ref{fig:joints} and Fig. \ref{fig:jointsGeometry} to be sent directly to the 2D cutting machine (e.g. laser cutter and waterjet). During this process, the fabrication specifications are considered, and joints are selected and rendered in the final files. A typical manufacturing specification is material. Fabrication parameters of machine will be determined on a case-basis resulting in a unique fabrication output file. Another example of fabrication specifications is the thickness of material. With different thicknesses, the joint pattern would be various from one to one (see Section \ref{sec:joints}). All of these fabrication specifications will finally result in variations in 2D fabrication pattern, such as laser cutting kerf and amplitude of interference. Our system also is able to render 3D models (.STL files, see Fig. \ref{fig:connectionrepr}A, \ref{fig:hierarchicalcomp}, and \ref{fig:connectiondetec}A for visualization to assist design process.

In some other design tools\cite{Schulz2014}, even the furniture can be designed with an ease, the fabrication and assembly processes still require a lot of experience, which keeps the users from designing and manufacturing their own furniture. Our solution for this issue is to adopt joints that are available for 2D cutting and easy to assemble as mentioned in Section \ref{sec:joints}. These planar joints are easy to align with due to the special configuration and simple to assemble through interference fit without the requirement for extra carpentry tools and skills. Our method is validated through several successful assemblies (see Fig. \ref{fig:furniturecollection}, Fig. \ref{fig:designManipulation}, and \ref{fig:compounddesign}) in the paper. 
\section{Results and Discussion}
\label{sec:resultnDiscussion}
To demonstrate the proposed system in this paper, a variety of (scaled) furniture are designed, manufactured, and assembled. In this section, in order to have a quick demonstration, we choose 3mm and 6mm plywood as raw materials to fabricate (scaled) furniture. The fabrication of real size furniture are very similar. We use a laser cutter to pattern these materials and assemble components manually through interference fit (see Section \ref{sec:joints}).  Full-sized furniture could be created by using a waterjet on thicker plywood; a similar assembly process could be used to assemble those pieces.

\subsection{Design Examples}
\begin{figure}[t]
    \center
    \includegraphics[trim=0cm 9.4cm 0cm 0cm, clip=true,width=1\textwidth]{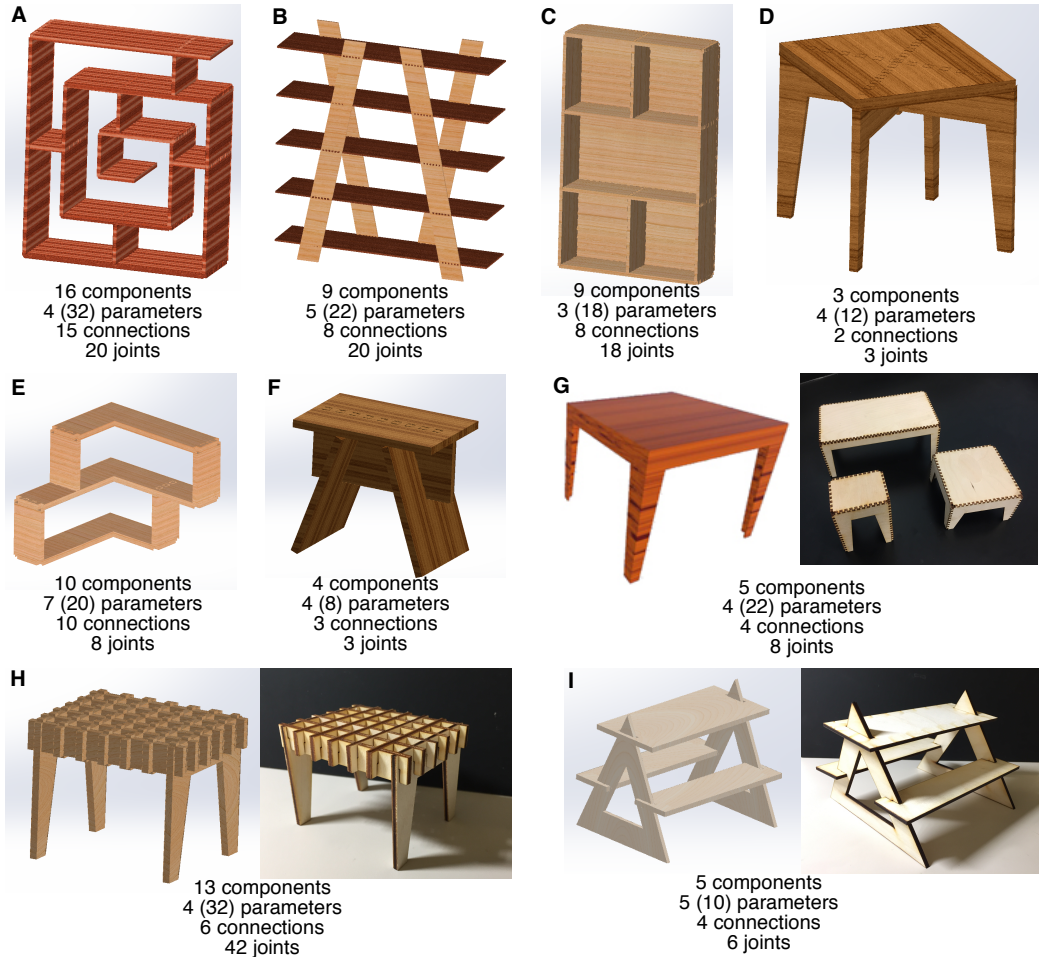}
    \caption{Furniture design examples using our system. The numbers of components, available design parameters, connections and joints of each design are labelled. For design parameters, the structure-preserved values are calculated after some constraints are added to preserve the structure of the designs while the maximal values are included in brackets.}
    \label{fig:furniturecollection}
\end{figure}

We start with a collection of rather simple furniture designs (see Fig. \ref{fig:furniturecollection}) by simply connecting existing elementary components together. These components are typically simple geometric faces, such as triangles, squares and polygons. In most cases, users only need to define a small number of connections, which can be far less than the number of joints. For example, users only need to specify 6 connections to define the spatial relations for all components of a stool in Fig. \ref{fig:furniturecollection}H. Our system will automatically place all 42 joints (slot-slot joints), which greatly lowers the difficulty of creating this furniture design. The same phenomenon can be found in most of the designs as shown in Fig. \ref{fig:furniturecollection}.

In addition, our intersection detection algorithm can help to merge coplanar components to reduce the number of components of a specific design, which lowers the difficulty of fabrication and assembly for causal end-users. For instance, in Fig. \ref{fig:furniturecollection}E, two trapezoids are combined into a L shape component due to coplanar faces merge, which means we have fewer components and joints. Also, the usage of built-in planar joints significantly reduces the number of constructing elements (in particular joints) for similar furniture designs \cite{Schulz2014} (also see Fig. \ref{fig:designManipulation} and \ref{fig:compounddesign}). This reduction can make the design, fabrication, and assembly processes much easier. Lastly, every model in our system has a large number of continuous design parameters, which allows users to freely customize their furniture to match their desires. After specific constraints added to the original design parameter set, a handful of meta-parameters are obtained to allows for design manipulation with structure preserved. For instance, the simple table in \ref{fig:furniturecollection}G, has 4 structure-preserved design parameters, including the length, width, height of the table, and the width of table leg. However, an unconstrained table can have up to 22 parameters allowed to be modified to generate a much broader set of designs. 

\subsection{Flexibility of Modeling Process}
\label{sec:modelingFlexibility}
\begin{figure}[t]
    \center
    \includegraphics[trim=0cm 23.8cm 0cm 0cm, clip=true,width=1\textwidth]{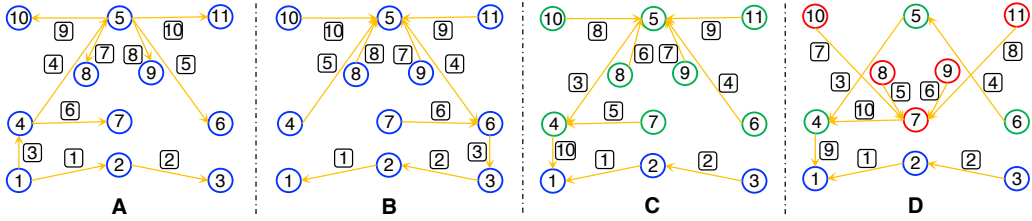}
    \caption{An illustration of the flexibility of modeling process in our system. Components labelled as circled number, e.g. \circled{1}) and connections represented by boxed number, e.g. \squared{1}. Components from different furniture models are differentiated by color. \textbf{A)} Opposite connection direction but the same ordering; \textbf{B)} Different connection ordering; \textbf{C)} Hierarchically composed from two different designs; \textbf{C)} Hierarchically composed from three models.}
    \label{fig:modelFlexibility}
\end{figure}

It is important that the modeling or designing process of a furniture is flexible and robust, which allows casual end-users to freely construct a furniture design as they prefer with any ordering or approach. Enabled by the abstract scheme and unique intersection algorithm (see Section \ref{sec:intersectiondete}), users can obtain the same desired model (along with fabrication file) with different design processes (e.g. different order or hierarchy). We use the reading desk (see Fig. \ref{fig:connectionrepr}A) as example to show how users can build a furniture differently. Users can follow the same connection order as the original design but flip the direction of every connection (see Fig. \ref{fig:modelFlexibility}A). For instance, connection \squared{1} is flipped from $\squared{1}  (\circled{2}\xrightarrow{}\circled{1})$ to $\squared{1}  (\circled{1}\xrightarrow{}\circled{2})$). Still, users obtain the same final model as previous. Users can use a more intuitive order of connections as they prefer. One of the example orders is shown in Fig. \ref{fig:modelFlexibility}B. In addition, as illustrated in Fig. \ref{fig:modelFlexibility}C, users are enabled to hierarchically compose the reading desk from two simple furniture models, such as a simple table (composed of component 1-3) and a trapezoid shelf (composed of component 4-11). Similarly, the desk can be composed from three low-level models (see Fig. \ref{fig:modelFlexibility}D). This high flexibility of modeling process allows users only focus on design itself instead of tedious engineering details.

\subsection{Design Manipulation}
\label{sec:manipulation}
\begin{figure}[t]
    \center
    \includegraphics[trim=0cm 15.6cm 0cm 0cm, clip=true,width=1\textwidth]{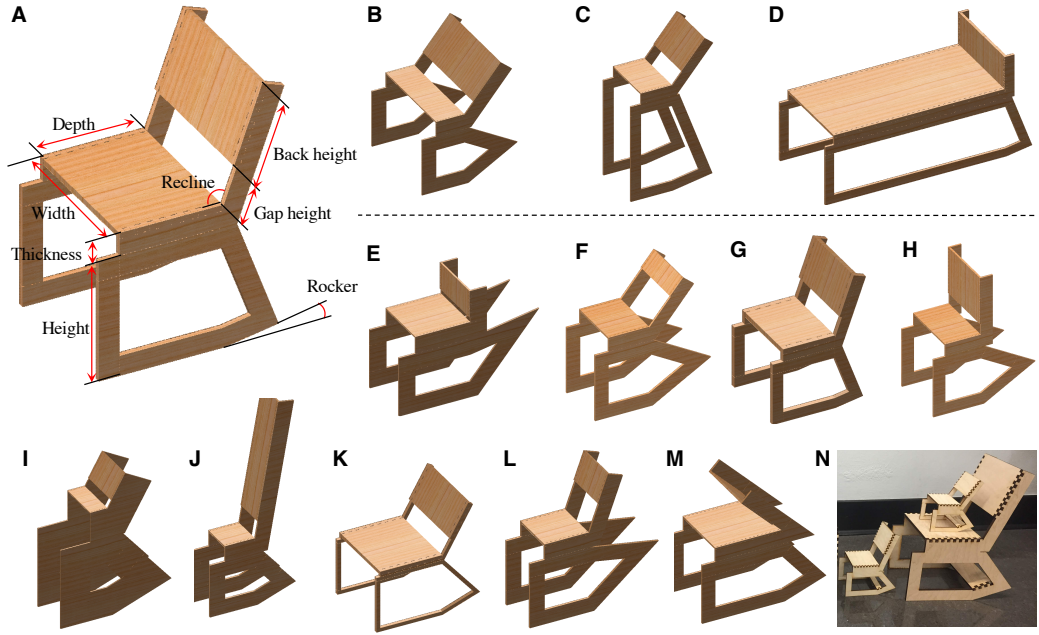}
    \caption{An illustration of design manipulation after a rocker chair has been composed in our system. 8 metaparameters of a rocker chair are labelled \textbf{A)}. A bunch of variants of the rocker chair are created by manipulating the metaparameters. These variants included a long bench for the side of the pool \textbf{B)}, a tall chair for bar \textbf{C)}, and a rocking bed \textbf{D)}. Other wildly modified rocker chairs are shown in \textbf{D)}-\textbf{M)} to demonstrate the vast design space. Several scaled chairs are fabricated and assembled \textbf{N)}.}
    \label{fig:designManipulation}
\end{figure}

Owing to the parameterized modeling scheme, users are allowed to manipulate parameters of a design, which means users are capable of modifying a furniture's dimensions while preserving its overall structure. Design manipulation is permitted at all levels of hierarchy. Therefore, the user can make higher-level modifications by editing the hierarchical composition and make more detailed changes by selecting low-level internal nodes. Fig.~\ref{fig:designManipulation} shows an example of how users can continue to explore the design space of a furniture model. For this rocker chair design, eight metaparameters are defined by adding some geometric constraints to greatly reduce the design parameters for structure-preserved manipulation. By varying these eight parameters, users can have a bunch of variants that have distinct functionalities for various applications. For example, users can widen the chair to obtain a long bench for the side of a pool, as shown in Fig.\ref{fig:designManipulation}B. Users are also allowed to modified the parameters wildly to create extreme designs, as shown in Fig.\ref{fig:designManipulation}E-M, for special applications. In addition, some scaled rocker chairs are fabricated and assembled to validate the design manipulation (see Fig.\ref{fig:designManipulation}N).

For users with more expertise, they can impose constraints between related parameters of its constituent parts. For example we can add constraints between two edges to make then always equal to each other to simplify the design process if they are supposed to be equal. Also, causal end-users can benefit from the constraints encoded by experts by narrowing down the parameters of designs, as the above demonstrated metaparameter definition.

\subsection{Compound Designs}
\begin{figure}[t]
    \center
    \includegraphics[trim=0cm 12.1cm 0cm 0cm, clip=true,width=1\textwidth]{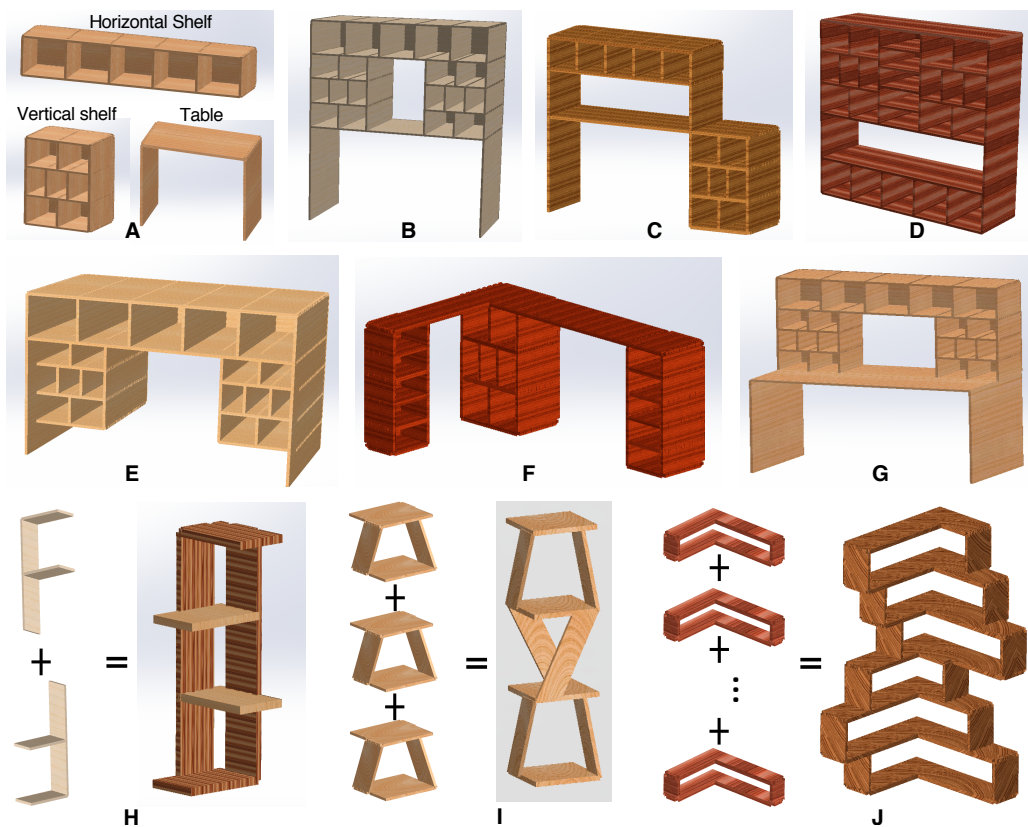}
    \caption{Compound furniture designs hierarchically composed from existing furniture models. \textbf{A)} A collection of simple furniture models, including a vertical shelf, a horizontal shelf, and a simple table. 6 compound furniture models, composed from the aforementioned simple designs are displayed from \textbf{B)} to \textbf{G)}. They are a TV console B), a study desk C), a multi-use shelf D), an over-the-toilet storage E), a corner workshop bench F), and a dresser G), respectively. Several other compound designs are also presented in \textbf{H)}, \textbf{I)}, and \textbf{J)}.}
    \label{fig:compounddesign}
\end{figure}

Though the users can always design their furniture from scratch, it is extremely challenging to do so for those who do not possess domain skills. Our hierarchical scheme will save them by composing complex furniture from simpler ones. This hierarchical implementation resolves the challenge by breaking down the complicated design process into recursively combining the relevant simpler building blocks. Fig. \ref{fig:compounddesign} demonstrates how users can compose several simple designs into more complicated compound furniture models. The building blocks are vertical shelves, horizontal shelves, and simple tables (see Fig. \ref{fig:compounddesign}A). The constructions of these building blocks are rather easy. By hierarchically combining these building blocks, we can obtain numerous various furniture models as shown in Fig. \ref{fig:compounddesign}B-G. Basically, users merely need to specify how these building blocks are placed against each other, leaving the tedious engineering implementations, such as joints intersection, to our system. Take a study desk as example (see \ref{fig:compounddesign}C). Firstly, users stack two tables vertically. Secondly, users add a horizontal shelf on the top of the previously composed design. Lastly, users attach a vertical shelf on the right sideboard of the bottom table to finalize the study desk design. Other compound designs can be created in a similar manner, which greatly releases causal end-users from the tedious work. As you can see, this hierarchical process decomposes the complex modeling into a series of trivial composition steps. The power of this hierarchical approach is also validated by other several hierarchically composed furniture designs shown in Fig. \ref{fig:compounddesign}H, I, and J. Combining several basic units can result in sophisticated designs. For example, as shown in Fig. \ref{fig:compounddesign}I, a fancy bookshelf can be obtained by stacking three identical stools.

\section{Conclusion and future work}
\label{sec:conclusion}
In this paper, we have proposed a computational, function-based design pipeline for digital fabrication of flat-pack furniture. Our system enables casual end-users to easily design, customize, fabricate and assemble furniture models by leveraging parameterized abstraction, hierarchical composition, intersection auto-detection, and planar joint design. Moreover, thanks to the template-free design scheme, no predefined models are needed, which further reduces the dependence of design on domain experts\cite{Schulz2014}. We have demonstrated the power of our approach by designing, fabricating, and assembling various (scaled) furniture models. Our method also shows the potential of introducing the design of customizable furniture to the general public by greatly reducing the required resources. In conclusion, we present a fully computational design tool for casual end-users to design their own flat-pack furniture that are guaranteed to be manufacturable and easy to assemble. We believe that our work, together with the flat-pack furniture library we release, will inspire interesting future studies. For instance, creating an user-friendly interface will be a natural extension of our work, which will further facilitate the design process.

Also, exploring the assemblability of resulting furniture models in our system is an interesting topic. Thanks to the configuration of current planar joints, the manufacturability of all designs is guaranteed. Although the assemblability of furniture models has not been validated theoretically, all of the created designs in this paper are verified manually to be assemblable. However, the assemblability of designs will be difficult to validate when models get much more complicated or more types of joint are involved. Thus, an algorithm for assemblability validation is necessary at that time. For now, we tackle this challenge by using low cost substitutions (e.g. copy paper) to build prototype for a specific design to check its assemblability of the original design. 

One of the most under-explored elements of our pipeline is the physical simulation. There are many mechanical properties of furniture designs that need to be investigated in order to guarantee their functionality. These properties include strength of joints, stability and stress distributions of furniture models under typical loading in daily life. Given the abstract and hierarchical scheme of our system, components and designs are frequently reused to construct more complex models, which suggests a data-driven approach to accelerate the physical simulations.

Lastly, it would be exciting to extend our work to enable active furniture design. Active designs will add reconfigurability to furniture, which allows more interactions with humans in everyday life and can enriches the design space for smart homes\cite{Wiki2018}. Since our framework is abstract and parameterized, it is straightforward to add new joints and active components to the system. The most challenging part is to define a set of appropriate joints to allow active movements. Thanks to our abstract design scheme, it is possible to replace the stationary planar joints with special active counterparts with controllable actuation (e.g. servo-driven actuation\cite{Mehta2015}). The active connections could be implemented by the flexible joint, as shown in Fig.\ref{fig:flexible_joint}, which allows angular movements. On the other hand, the ability to codesign electrical components and controlling software for actuation is another challenge, which can refer to the method proposed by Mehta et al.\cite{Mehta2014}. One example of this active design can refer to Fig.\ref{fig:joints}E-G. 

Ultimately, we have presented an abstract and hierarchical approach as a very general method for design, fabrication and assembly, which lays the foundation of exploring this class of flat-pack furniture. In the future, we believe this method will enable causal end-users to design, manufacture and assemble various models, such as furniture, architecture, robots, and beyond.

\section*{Acknowledgements}
The authors would like to thank Mr. Christian Warloe and Mr. Gopi Suresh for their valuable comments and helpful suggestions.
  
This work is supported by the National Science Foundation under grant \#1644579: A Comptuational Approach to Customizing Design, for which the authors express thanks.


\appendix
\section{Appendix}
\subsection{Current Joint Collection}
\label{sec:joints}
In this section, we introduce the details about three planar joints implemented in our paper. 
\subsubsection{Finger-Finger Joint}
\label{sec:finger_finger}
\begin{figure}[h]
    \center
    \includegraphics[trim=0cm 23.6cm 0cm 0cm, clip=true,width=1\textwidth]{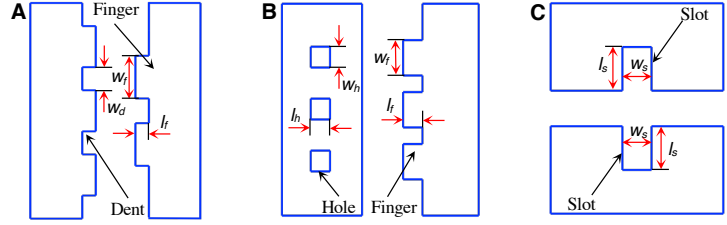}
    \caption{A collection of planar joints used in our system. \textbf{A)} Finger-finger joints; \textbf{B)} Finger-hole joints; \textbf{C)} Slot-slot joints.}
    \label{fig:jointsGeometry}
\end{figure}

\hfill \break
Finger-finger joints are used to connect two components with their segment of intersection both on the edges. The geometric outline of this finger-finger joint is shown in Fig. \ref{fig:jointsGeometry}A, where alternating rectangular fingers are added to the two coupled edges, and the length of these fingers $l_f$ are matched with the thickness of the material (3mm in this case) to ensure that the seamless outline of the attached edges. The interval between two fingers forms a dent, which is fulfilled by a finger when the joint is assembled.

As shown in Fig. \ref{fig:jointsGeometry}A, to ensure that the finger-finger joint can indeed hold components together, we introduce a certain amplitude of interference to the joint. In addition, a geometry correction is necessary to compensate for the fabrication kerf introduced by the machines. Thus, when user has specified a interference fit value $\delta$ and a certain fabrication method, the convex finger is expanded to the designated value $w_f$, while the concave dent is trimmed down accordingly. The relationship between these two parameters is described by: $$w_f= w_d + 4\Delta + \delta$$ where $\Delta$ is the fabrication kerf and $\delta$ is the interference amplitude. For the specific material and fabrication method used, users may need to perform some experiments to find the best interference fit value and the fabrication kerf.

Due to the limitation of 2D fabrication, finger-finger joints are only well suited for connections at angle of 90\textsuperscript{$\circ$}. Although user can easily spin components around to form other angles of connections, we do not regard it as the expected usage of finger-finger joint due to the questionable firmness of such connections. 

\subsubsection{Finger-Hole Joint}
\label{sec:finger_hole}
\hfill \break
If the segment of intersection is on the edge of one component while within the face of the other, finger-hole joints are needed. We add fingers to the former component and holes to the latter component (see Fig. \ref{fig:jointsGeometry}B). Similar to finger-finger joints, users can specify the interference fit value to determine the final geometry of the joint. The relationship between the widths of finger $w_f$ and hole $w_h$ is expressed by: $$w_f= w_h + 4\Delta + \delta$$ where $\Delta$ is fabrication kerf and $\delta$ is interference amplitude. Again the length $l_f$ of these fingers and length $l_h$ of holes should both be equal to the thickness of the material. Finger-hole joints are also limited to 90\textsuperscript{$\circ$} connections.

\subsubsection{Slot-Slot Joint} 
\label{sec:slot_slot}
\hfill \break
If the segment of intersection is within the face of both components, we will add a slot-slot joint to connect them. If neither face is fully within each other, then we will cut rectangular slots on both components, and each slot accounts for half of the segment of intersection (see Fig. \ref{fig:jointsGeometry}C). If one face is fully with another, then we will only cut the slot on one face so that the other face can be stuck through it(see Fig. \ref{fig:furniturecollection}I). 

The length of the slot $l_s$ is half of the length of the segment of intersection and the width $w_s$ of the slot can be expressed by: $$w_s=w_m + 2\Delta + \delta$$ where $w_m$ is the thickness of material, $\Delta$ is the fabrication kerf and $\delta$ is the interference amplitude.

\subsection{Flexible Joint}
\label{sec:flexiblejoint}
\begin{figure}[t]
    \center
    \includegraphics[trim=0cm 21.5cm 0cm 0cm, clip=true,width=1\textwidth]{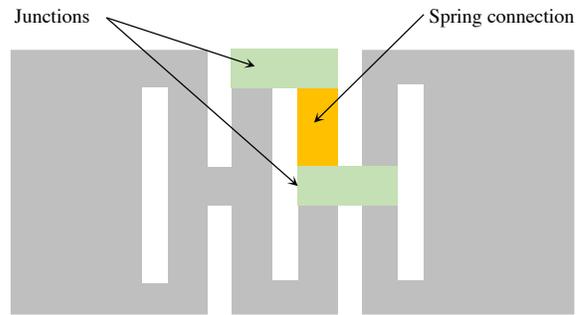}
    \caption{A basic module of a flexible joint, with its junctions and spring connection structure labelled. The junctions are supposed to behave as rigid joints while the spring connections are responsible for plate bending.}
    \label{fig:flexible_joint}
\end{figure}

A basic module of flexible joints is shown in Fig. \ref{fig:flexible_joint}. For wider flexible joints, this pattern could be replicated several times along the transverse direction. This module consists of junctions and spring connections. The junctions behave as rigid joints to connect spring connections. The spring junctions function as bending spring to enable angular movement. More details about the design of flexible joints can be found in the reference\cite{Blashki2015}. 

\end{document}